\documentstyle[psfig]{elsart}

\def\lsim{\mathrel{\rlap{\lower4pt\hbox{\hskip1pt$\mathchar"0218$}}
       \raise1pt\hbox{$<$}}}
\def\gsim{\mathrel{\rlap{\lower4pt\hbox{\hskip1pt$\mathchar"0218$}}
       \raise1pt\hbox{$>$}}}

\begin{document}
\begin{frontmatter}
\title{Charge Stripes and Antiferromagnetism in\\
Insulating Nickelates and Superconducting Cuprates}
\author{J. M. Tranquada}
\address{Physics Department, Brookhaven National Laboratory, Upton, NY
11973, USA}

\begin{abstract}
Neutron and X-ray scattering studies have provided strong evidence for coupled
spatial modulations of charge and spin densities in layered nickelates and
cuprates.  The accumulated results for La$_{2-x}$Sr$_x$NiO$_{4+\delta}$ are
consistent with the strongly-modulated topological-stripe concept.  Clues from
Nd-doped La$_{2-x}$Sr$_x$CuO$_4$ suggest similar behavior for the cuprates.  The
experimental results are summarized, and features that conflict with an
interpretation based on a Fermi-surface instability are emphasized.  A
rationalization for the differences in transport properties between the
cuprates and nickelates is given.
\end{abstract}
\end{frontmatter}

\section{Introduction}

Both La$_2$NiO$_4$ and La$_2$CuO$_4$ are two-dimensional antiferromagnetic
insulators.  Their insulating natures are due to the strong Coulomb repulsion $U$
between electrons occupying the same transition-metal $3d$ orbital.  As recently
emphasized by Anderson \cite{ande97}, a sufficiently large $U$ can cause the
$3d$ electrons to be localized.  Virtual hopping of the electrons between
neighboring metal sites lowers their kinetic energy in second order
perturbation theory, and when combined with the Pauli exclusion principle, this
leads to the antiparallel alignment of neighboring spins.

Various spectroscopic studies have confirmed that $U$ is indeed sufficiently
large to cause strong correlations.  The fact that the large insulating gaps
remain when samples are heated above their respective N\'eel temperatures
further demonstrates that these materials are not Slater antiferromagnets or
spin-density-wave systems.  Thus, the appropriate starting point for
considering the insulating nickelate and cuprate compounds consists of
localized electrons with effective interactions via superexchange.

What we are really interested in, however, is what happens when holes are doped
into the $M$O$_2$ planes ($M=$ Cu or Ni).  In the case of the cuprate, a hole
concentration $n_h\gsim0.05$ per Cu  yields an unusual metal that becomes
superconducting at surprisingly high temperatures.  The nickelate, on the other
hand, remains insulating even when $n_h$ is increased to 0.5.  Such strongly
contrasting behavior might lead one to expect that the electronic correlations
in these systems evolve in quite different manners; however, neutron scattering
studies have shown that the magnetic (and possibly charge) correlations follow
very similar trends with doping.  In this paper I will review these trends, and
consider the relevance of various theoretical approaches for interpreting them.

\section{La$_{2-x}$Sr$_x$NiO$_{4+\delta}$}

Consider first the nickelates.
At low doping it appears that the holes must be localized near the dopants.  As
a result, the details of the phase diagram in this regime depend on the nature
of the dopant, substituted Sr or excess oxygen.  Once the hole concentration is
large enough, $n_h\gsim0.2$, coupled charge- and spin-density modulations start
to form within the NiO$_2$ planes.  These modulations align themselves diagonally
with respect to the Ni-O bonds, as indicated by the positions of the
charge-order and magnetic superlattice peaks observed by electron \cite{chen93}
and neutron diffraction \cite{tran95b,tran96a,lee97}.   If one makes use of a
unit cell with axes rotated by
$45^\circ$ with respect to the Ni-O bonds, then in reciprocal space the magnetic
peaks are split about the antiferromagnetic position (1,0,0) along the [100] and
[010] directions by an amount $\epsilon$ (in reciprocal lattice units), while the
charge-order peaks are split about fundamental Bragg peaks by $2\epsilon$. 
(Neutrons do not scatter from charge directly, but instead are sensitive to the
atomic displacements induced by the charge-density modulation.  The 
charge-order peaks have also been observed by X-ray diffraction \cite{vigl97}.)
The relationship between charge and spin wave vectors indicates that the period
of the spin structure in real space is twice that of the charge modulation. 
Measurements of higher-order magnetic harmonics \cite{woch97} and the magnitude
of magnetic moments \cite{tran95b} are consistent with a strong localization of
the charge into stripes of the type suggested by calculations of Zaanen and
Littlewood \cite{zaan94}.  In this picture, the regions between stripes are
antiferromagnetic domains, with a phase that shifts by $\pi$ on crossing a
stripe.  Experiments on a range of samples show that
$\epsilon\approx n_h=x+2\delta$ at low temperature.  This indicates that adding
more holes results in more stripes of roughly constant charge density (roughly
1 hole/Ni) which must move closer together.

The relationship between charge-order and magnetic-order wave vectors is just
the same as that found in Cr \cite{pynn76}.  In the latter system, the dominant
spin-density wave (SDW) order is consistent with Fermi-surface nesting.  The
weak charge order appears at the same temperature as the SDW, and the CDW order
parameter varies as the square of that for the SDW.  In the nickelates, the
situation is rather different.  The charge order always appears at a higher
temperature than the magnetic order, and both ordering temperatures tend to
increase linearly with $n_h$.  Such behavior is inconsistent with order driven
purely by the spin density \cite{zach97}.  Furthermore, optical reflectivity
measurements on a Sr-doped sample with $x=\frac13$ indicate that the NiO$_2$
planes do not become metallic even at temperatures well above the ordering
temperature \cite{kats96}.

\begin{figure}
\centerline{\psfig{figure=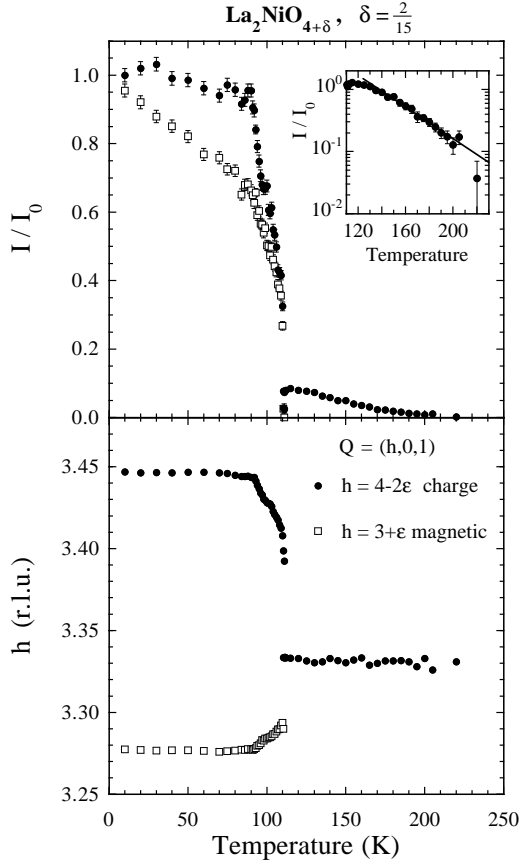,width=7cm}}
\caption{Temperature dependences of a magnetic peak at $(3+\epsilon,0,1)$ (open
squares) and a charge order peak at $(4-2\epsilon,0,1)$ (filled squares)
measured by neutron diffraction on a single crystal of La$_2$NiO$_{4.133}$. 
Upper panel shows integrated intensities normalized at 10~K; lower gives the
peak position in $h$.  From Wochner {\it et al.} \protect\cite{woch97}.}
\end{figure}

Other relevant results have been obtained in a recent neutron diffraction study
of La$_2$NiO$_{4+\delta}$ with $\delta=\frac2{15}$ \cite{woch97}.  As shown in
the lower panel of Fig.~1, for this sample $\epsilon$ is temperature dependent;
more precise measurements of a third-harmonic peak position reveal lock-in
plateaus at certain rational fractions.  In the case of Fermi-surface-driven
order, one would expect the wave vector to lock in at a temperature-independent
value, $2k_F$.  In contrast, the observed behavior suggests that competing
interactions are involved, one of which is likely to be the long-range part of
the Coulomb interaction.

\section{La$_{2-x}$Sr$_x$CuO$_4$}

Evidence for incommensurate magnetic fluctuations in La$_{2-x}$Sr$_x$CuO$_4$ was
first provided by the inelastic neutron scattering measurements of Cheong {\it
et al.} \cite{cheo91}.  To describe this it is convenient to express results
in terms of a unit cell with axes parallel to the Cu-O bonds, so that the
antiferromagnetic wave vector is $(\frac12,\frac12,0)$.  The magnetic
scattering is peaked at positions split about this wave vector by an amount
$\epsilon$ along [100] and [010] directions.  Yamada {\it et al.}
\cite{yama97b} have shown that $\epsilon\approx x$ for $0.05\lsim x\lsim\frac18$,
and $\epsilon\approx\frac18$ for $x\gsim\frac18$.

The fact that La$_{2-x}$Sr$_x$CuO$_4$ is metallic has caused some researchers
to look to Cr as a model for understanding the incommensurate magnetic
scattering \cite{maso92}.  While it is true that Cr alloys exhibit similar
magnetic scattering that can be purely inelastic or have an $E=0$ component
\cite{fawc94}, these alloys always tend to be good metals, and it is not
possible to achieve a correlated insulator state by doping.  In contrast, many
of the experimentally observed features in the cuprates, such as the lack of a
shift in the chemical potential in La$_{2-x}$Sr$_x$CuO$_4$ for $x\lsim0.15$
\cite{ino97}, bear strong similarities to doped semiconductors; hence, proximity
to an insulating phase appears to be an important feature of the phase diagram. 
Recently it has been argued that a generic feature of 2D doped insulators is the
generation of topological defects in the form of antiphase domain walls
\cite{kive96,zaan97}.  In the case of doped La$_2$NiO$_4$, we have seen that
these topological stripes have scattering signatures similar to those of Cr
alloys even though the underlying physics appears to be different.

To gain further clues on the nature of the magnetic correlations in the
cuprates, studies have been performed on La$_{1.6-x}$Nd$_{0.4}$Sr$_x$CuO$_4$. 
The substituted Nd has the same valence but a smaller ionic radius than the La,
and the latter feature induces a subtle change in the low-temperature tilt
pattern of the CuO$_6$ octahedra \cite{craw91,buch94a}.  Associated with this
structural modification is a strong suppression of $T_c$ for
$x\approx\frac18$.  Neutron diffraction studies \cite{tran95a,tran96b} on a
single crystal with $x=0.12$ have shown the appearance of {\it elastic} magnetic
scattering with the same incommensurate splitting $\epsilon$ as that found in
crystals with the same $x$ and no Nd \cite{yama97b}.  Furthermore, weak peaks
consistent with charge order have been found with neutrons
\cite{tran95a,tran96b} and confirmed with high-energy X-ray scattering
\cite{vonz97}.  The charge order appears at a higher temperature than the
magnetic order, as in the nickelates.  Further work has shown that the static
magnetic order coexists with superconductivity in Nd-doped crystals with
$x=0.15$ and 0.20 \cite{tran97a,oste97}.

At first glance, one might guess that the depression of $T_c$ in the Nd-doped
crystals might be due to a change in the density of states near $E_F$
associated with the modification of the crystal structure; however, band
calculations based on experimentally determined crystal structures indicate
that the phase change has relatively little effect on the density of states
\cite{norm93}.  If the magnetic and charge order were driven by a Fermi surface
instability, it would have to be a CDW-type instability because of the distinct
ordering temperatures \cite{zach97}; however, the wave vector for the charge
order ($2\epsilon=0.24$ rlu) is not a special spanning vector of the Fermi
surface.  Also, the photoemission measurements on La$_{2-x}$Sr$_x$CuO$_4$
reported by Fujimori {\it et al.} \cite{fuji97} do not provide any evidence for
coherent quasiparticles near $E_F$; such quasiparticles would be a prerequisite
for any Fermi-surface-driven ordering.  Finally, optical reflectivity
measurements on Nd-doped crystals \cite{taji97} indicate that the
charge-excitation gap one  would expect in the case of a CDW would have to be
exceedingly small.  

If the spin and charge modulations are related to topological stripes, then one
would expect the amplitudes of the modulations to be large.  Although direct
measurement of the amplitude of the charge modulation is difficult, local-probe
measurements of the spin amplitude are available \cite{luke97,wage97,roep97}. 
In particular, muon spin rotation ($\mu$SR) studies on samples of
La$_{2-x-y}$Nd$_y$Sr$_x$CuO$_4$, with $x=0.12$, $y=0.4$ \cite{luke97}, and with
$x=0.15$, $y=0.3$--0.6 \cite{wage97}, indicate a maximum ordered moment
approximately half that in La$_2$CuO$_4$, and an ordering temperature of
$\sim30$~K.  The magnitude of the ordered moments is substantially
reduced compared to the antiferromagnetic insulator state, but at the same time,
the ordering temperature is diminished by an even greater amount.  One might
expect quantum spin fluctuations to affect both quantities, but how can we
evaluate such effects?  Castro Neto and Hone \cite{neto96} have proposed that
the main effect of the stripes is to reduce the exchange coupling between
neighboring magnetic domains, which can be modelled in terms of a spatially
anisotropic nonlinear sigma model.  Using such a model, van Duin and Zaanen
\cite{vand97} have calculated the ordering temperature $T_N$ (which requires an
assumption about interlayer coupling) and low-temperature staggered
magnetization $M_s$ as a function of anisotropy, $\alpha$, where $\alpha=1$
corresponds to isotropic interactions and the local superexchange is held
constant.  Although the magnitude of the effective anisotropy is not known, a
useful way to present the results is to plot the variation of $T_N(\alpha)$
versus $M_s(\alpha)$, as shown in Fig.~2.  Normalizing the experimental results
on the modulated phase to those for La$_2$CuO$_4$, we see that they are fairly
consistent with this model.  Thus, the observed spin-density amplitude is
compatible with the strong modulation expected for topological stripes.

\begin{figure}
\centerline{\psfig{figure=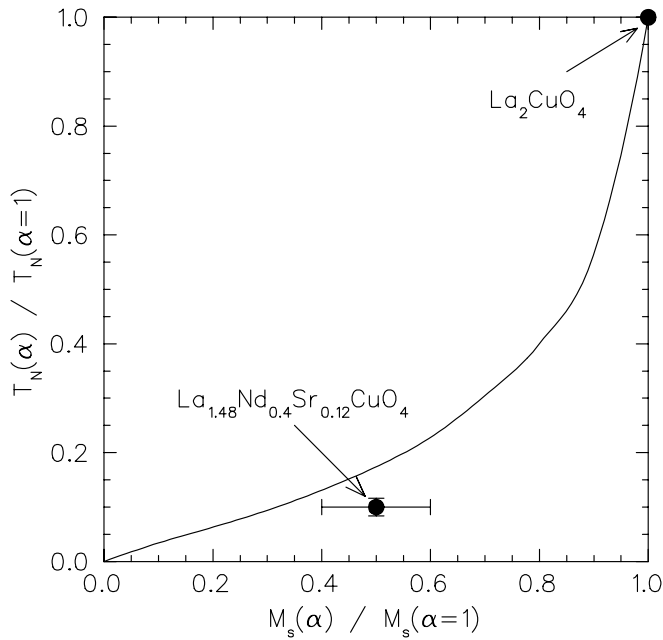,height=3.3in}}
\caption{Solid line: theoretical variation
of the relative N\'eel temperature vs.\ low-temperature
staggered magnetization as an implicit function of
exchange anisotropy associated with stripes.  From a model
calculation by van Duin and Zaanen \protect\cite{vand97}.  The point labelled
La$_{1.48}$Nd$_{0.4}$Sr$_{0.12}$CuO$_4$ is from a $\mu$SR experiment by Luke
{\it et al.} \protect\cite{luke97}.	}
\end{figure}

The simplest caricature of ordered charge stripes involves a non-sinusoidal
modulation of the charge and spin densities.  The stripe interpretation has been
questioned due to the lack of evidence for strong higher-harmonic superlattice
peaks that might be expected.  It turns out, however, that such a concern is
not relevant.  First of all, as demonstrated in the case of a nickelate sample
\cite{woch97}, the intensities of higher harmonics corresponding to narrow
stripes are actually quite weak and would be difficult to observe in the
cuprates.  Secondly, fluctuations of stripes about their average positions
should smear the average distribution, leading to a further reduction in
harmonic intensities.  Thus, the absence of harmonics is not a valid
objection to stripes, but at the same time, models with sinusoidal
modulations, such as those discussed by Scalapino \cite{scal97}, are also
compatible with experiment. 

\section{Dynamic Correlations}

For a stripe-ordered phase, the low-energy spin fluctuations have a simple
interpretation: they are just the spin waves that one expects in a magnetically
ordered state.  Inelastic neutron scattering measurements on
La$_2$NiO$_{4.133}$ appear consistent with this concept \cite{tran97c}.  The
effective spin-wave velocity is reduced by about 40\%\ compared to the undoped
phase, where the velocity is proportional to the superexchange $J$ times the
lattice parameter $a$.  The softening of the spin waves in the stripe-ordered
phase is consistent with a reduction in the effective value of $J$ due to
relatively weak exchange coupling across domain walls
\cite{neto96}; however, the anisotropy that one would expect to see in the
spin-wave velocity for propagation parallel and perpendicular to the stripes has
not been clearly observed.  A quantitative test requires more detailed
theoretical predictions.

The important feature here is that the observed spin fluctuations are
associated with the magnetic moments in the antiferromagnetic domains, but not
directly with the spins of the doped holes.  When static stripe order
disappears, one would expect dynamic correlations to survive.  Low-energy spin
fluctuations should become overdamped, while modes at high enough energy
(sufficiently short wavelength) would be relatively insensitive to the loss of
order \cite{tran97d}.  For La$_{2-x}$Sr$_x$CuO$_4$ with $x=0.14$, the
effective spin-wave velocity associated with high-energy fluctuations appears
to be reduced compared to the pure antiferromagnet \cite{hayd96a}, as occurs in
the stripe-ordered nickelate discussed above.  In contrast, if the spin
fluctuations were related to the response of itinerant electrons near the Fermi
surface, the velocity should be comparable to the Fermi velocity $v_F$ instead of
$Ja/\hbar$.  Band-structure calculations \cite{alle87} indicate that $v_F$ is 2
to 3 times the spin-wave velocity of La$_2$CuO$_4$ \cite{aepp89}.

In La$_{2-x}$Sr$_x$CuO$_4$ with $x=0.14$, where no static stripe order is
observed at finite temperature, the dynamics of the magnetic fluctuations might
exhibit quantum critical behavior.  In fact, evidence for such behavior is
reported by Aeppli {\it et al.} \cite{aepp97}.  A distinct quantum critical
point, associated with an incommensurate CDW instability, has been discussed by
Di Castro and coworkers \cite{dica97}.

Of course, any correlations that are relevant to the superconductivity must be
common to all of the superconducting cuprates.  For this reason, the recent
observations of incommensurate magnetic scattering in YBa$_2$Cu$_3$O$_{6.6}$,
discussed by Mook \cite{mook97}, are quite important.  Connections suggested by
the response to Zn doping have been listed elsewhere \cite{tran97e}.  
As an aside to the question of universality, it is interesting to note that
stripe modulations have also been observed in electron diffraction studies of
doped manganites \cite{cheo97}.

\section{Differences Between Cuprates and Nickelates}

I have argued that there are strong similarities between the spatial correlations
of charge and spin in the cuprates and nickelates; however, as mentioned in the
introduction, the nickelates remain insulating at dopant concentrations where
the cuprates become metallic and superconducting.  If the hole concentration in
the nickelate stripes were precisely 1 hole/ Ni site, then this would be
equivalent to a half-filled 1D system.  Such a system should be a Mott
insulator \cite{ande97}.  However, this explanation is incomplete in the case
of La$_2$NiO$_{4.133}$ where it has been demonstrated that the charge density
within the stripes must vary substantially with temperature
\cite{woch97}.  The deviation from $\frac12$-filling in these nickelate stripes
makes them similar to the cuprate stripes, which, when not pinned to the
lattice, apparently participate in metallic transport at finite temperatures. 
Thus it may not be possible to understand the differences between the
nickelates and cuprates simply in terms of the conductivity along a stripe.  An
alternative approach is to consider the motion of charge in the direction
transverse to the stripes.

Let us consider a 1D model for a charge stripe between antiferromagnetic
domains.  The cuprate case is shown at the top of Fig.~3, with the hole
centered on an oxygen atom.  The coupling between the O hole spin and the
neighboring Cu spins is responsible for the antiphase relationship between the
magnetic domains.  (A Cu-centered domain wall could be described as a
superposition of two such O-centered stripes.)  If the hole on O and a
neighboring Cu hole both shift one position to the left, the resulting
configuration is equivalent to the original one.  Thus, in this simple picture,
the 1D stripe would be expected to be itinerant.

\begin{figure}
\centerline{\psfig{figure=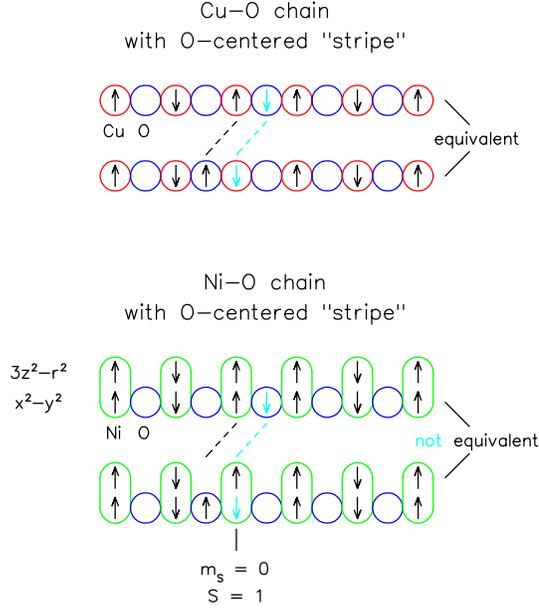,width=7cm}}
\caption{1D models of a topological charge stripe for Cu-O (top) and Ni-O
(bottom), as discussed in the text.}
\end{figure}

The situation changes in the nickelate case (bottom of Fig.~3) because of the
second $3d$ hole on each Ni.  The two $3d$ spins per Ni couple in a high-spin
triplet state.  Now when the O hole plus a $3d_{x^2-y^2}$ hole shift to the
left, we end up with a Ni site in a $m_s=0$, $S=1$ state \cite{zaanpc}.  This
shifted configuration is {\it not} equivalent to the original.  The shifted
state is higher in energy, and hence the stripe will tend to be confined in its
original position.  This effect should not depend on long-range order, and
hence may be relevant even when charge stripes become thermally disordered.

\section{Acknowledgments}

I have benefited from interactions with many experimental collaborators,
including J. D. Axe, D. J. Buttrey, G. Shirane, N. Ichikawa, S. Uchida, and P.
Wochner.  I am also grateful for frequent discussions with V. J. Emery and S.
A. Kivelson. This work is supported by Contract No.\ DE-AC02-76CH00016, Division
of Materials Sciences, U.S. Department of Energy.


\begin{thebibliography}{10}

\bibitem{ande97}
P.~W. Anderson, Adv. Phys. {\bf 46},  3  (1997).

\bibitem{chen93}
C.~H. Chen, S.-W. Cheong, and A.~S. Cooper, Phys. Rev. Lett. {\bf 71},  2461
  (1993).

\bibitem{tran95b}
J.~M. Tranquada, J.~E. Lorenzo, D.~J. Buttrey, and V. Sachan, Phys. Rev. B {\bf
  52},  3581  (1995).

\bibitem{tran96a}
J.~M. Tranquada, D.~J. Buttrey, and V. Sachan, Phys. Rev. B {\bf 54},  12318
  (1996).

\bibitem{lee97}
S.-H. Lee and S.-W. Cheong, Phys. Rev. Lett. {\bf 79},  2514  (1997).

\bibitem{vigl97}
A. Vigliante {\it et~al.}, Phys. Rev. B {\bf 56},  8248  (1997).

\bibitem{woch97}
P. Wochner, J.~M. Tranquada, D.~J. Buttrey, and V. Sachan, \null Phys. Rev. B
  (in press); cond-mat/9706261.

\bibitem{zaan94}
J. Zaanen and P.~B. Littlewood, Phys. Rev. B {\bf 50},  7222  (1994).

\bibitem{pynn76}
R. Pynn, W. Press, S.~M. Shapiro, and S.~A. Werner, Phys. Rev. B {\bf 13},  295
   (1976).

\bibitem{zach97}
O. Zachar, V.~J. Emery, and S.~A. Kivelson, \null Phys. Rev. B (in press).

\bibitem{kats96}
T. Katsufuji {\it et~al.}, Phys. Rev. B {\bf 54},  R14230  (1996).

\bibitem{cheo91}
S.-W. Cheong {\it et~al.}, Phys. Rev. Lett. {\bf 67},  1791  (1991).

\bibitem{yama97b}
K. Yamada {\it et~al.}, Physica C {\bf 282--287},  85  (1997).

\bibitem{maso92}
T.~E. Mason, G. Aeppli, and H.~A. Mook, Phys. Rev. Lett. {\bf 68},  1414
  (1992).

\bibitem{fawc94}
E. Fawcett {\it et~al.}, Rev. Mod. Phys. {\bf 66},  25  (1994).

\bibitem{ino97}
A. Ino {\it et~al.}, Phys. Rev. Lett. {\bf 79},  2101  (1997).

\bibitem{kive96}
S.~A. Kivelson and V.~J. Emery, Synth. Met. {\bf 80},  151  (1996).

\bibitem{zaan97}
J. Zaanen, \null J. Phys. Chem. Solids, this conference; cond-mat/9711009.

\bibitem{craw91}
M.~K. Crawford {\it et~al.}, Phys. Rev. B {\bf 44},  7749  (1991).

\bibitem{buch94a}
B. B\"uchner, M. Breuer, A. Freimuth, and A.~P. Kampf, Phys. Rev. Lett. {\bf
  73},  1841  (1994).

\bibitem{tran95a}
J.~M. Tranquada {\it et~al.}, Nature {\bf 375},  561  (1995).

\bibitem{tran96b}
J.~M. Tranquada {\it et~al.}, Phys. Rev. B {\bf 54},  7489  (1996).

\bibitem{vonz97}
M. von Zimmermann {\it et~al.}, (preprint).

\bibitem{tran97a}
J.~M. Tranquada {\it et~al.}, Phys. Rev. Lett. {\bf 78},  338  (1997).

\bibitem{oste97}
J.~E. Ostenson {\it et~al.}, Phys. Rev. B {\bf 56},  2820  (1997).

\bibitem{norm93}
M.~R. Norman, G.~J. McMullan, D.~L. Novikov, and A.~J. Freeman, Phys. Rev. B
  {\bf 48},  9935  (1993).

\bibitem{fuji97}
A. Fujimori {\it et~al.}, \null J. Phys. Chem. Solids, this conference.

\bibitem{taji97}
S. Tajima {\it et~al.}, \null J. Phys. Chem. Solids, this conference.

\bibitem{luke97}
G.~M. Luke {\it et~al.}, Hyp. Int. {\bf 105},  113  (1997).

\bibitem{wage97}
W. Wagener {\it et~al.}, Phys. Rev. B {\bf 55},  R14 761  (1997).

\bibitem{roep97}
M. Roepke {\it et~al.}, \null J. Phys. Chem. Solids, this conference.

\bibitem{neto96}
A.~H. {Castro Neto} and D. Hone, Phys. Rev. Lett. {\bf 76},  2165  (1996).

\bibitem{vand97}
C.~N.~A. van Duin and J. Zaanen, cond-mat/9707195.

\bibitem{scal97}
D.~J. Scalapino, \null J. Phys. Chem. Solids, this conference.

\bibitem{tran97c}
J.~M. Tranquada, P. Wochner, and D.~J. Buttrey, Phys. Rev. Lett. {\bf 79},
  2133  (1997).

\bibitem{tran97d}
J.~M. Tranquada, Physica C {\bf 282--287},  166  (1997).

\bibitem{hayd96a}
S.~M. Hayden {\it et~al.}, Phys. Rev. Lett. {\bf 76},  1344  (1996).

\bibitem{alle87}
P.~B. Allen, W.~E. Pickett, and H. Krakauer, Phys. Rev. B {\bf 36},  3926
  (1987).

\bibitem{aepp89}
G. Aeppli {\it et~al.}, Phys. Rev. Lett. {\bf 62},  2052  (1989).

\bibitem{aepp97}
G. Aeppli {\it et~al.}, \null Science (in press).

\bibitem{dica97}
C. {Di Castro}, \null J. Phys. Chem. Solids, this conference.

\bibitem{mook97}
H.~A. Mook, \null J. Phys. Chem. Solids, this conference.

\bibitem{tran97e}
J.~M. Tranquada, \null Physica B (in press); cond-mat/9709325.

\bibitem{cheo97}
S.-W. Cheong, \null J. Phys. Chem. Solids, this conference.

\bibitem{zaanpc}
J. Zaanen, (private communication).

\end{thebibliography}

\end{document}